\documentclass[amsmath,amssymb,aps,pra,10pt,notitlepage,groupedaddress,nofootinbib,longbibliography,]{revtex4-1}

\usepackage[utf8]{inputenc}
\usepackage{amsmath}
\usepackage{amsthm}
\usepackage{blindtext}
\usepackage{amssymb}
\usepackage{natbib}
\usepackage{graphicx}
\usepackage{color}
\usepackage{braket}
\usepackage[hyperfootnotes=true,colorlinks=true,allcolors=blue]{hyperref}
\usepackage{listingsutf8}
\definecolor{codegreen}{rgb}{0,0.6,0}
\definecolor{codegray}{rgb}{0.5,0.5,0.5}
\definecolor{codepurple}{rgb}{0.58,0,0.82}
\definecolor{backcolour}{rgb}{1, 1, 1}

\lstdefinestyle{codeblock}{
  backgroundcolor=\color{backcolour},
  commentstyle=\color{codegreen},
  keywordstyle=\color{blue},
  numberstyle=\tiny\color{codegray},
  stringstyle=\color{codepurple},
  basicstyle=\footnotesize,
  escapechar=\¢, 
  otherkeywords={with},
  breakatwhitespace=false,
  breaklines=true,
  captionpos=b,
  keepspaces=true,
  language=Python,
  numbersep=5pt,
  showspaces=false,
  showstringspaces=false,
  showtabs=false,
  tabsize=2,
  basicstyle=\ttfamily\footnotesize,
  inputencoding=utf8,
  upquote=true,
}
\usepackage{bbm}
\usepackage{bm}

\theoremstyle{plain}

\newtheorem{defn}{Definition}
\newtheorem{corollary}[defn]{Corollary}

\begin{document}

\title{Quantum computing with differentiable quantum transforms}

\author{Olivia Di Matteo\footnote{Current address: Dept. of Electrical and
  Computer Engineering, The University of British Columbia, Vancouver, BC, V6T
  1Z4, Canada}}
\author{Josh Izaac}
\author{Tom Bromley}
\author{Anthony Hayes}
\author{Christina Lee}
\author{Maria Schuld}
\author {Antal Sz\'ava}
\author{Chase Roberts}
\author{Nathan Killoran}
\affiliation{Xanadu, Toronto, ON, M5G 2C8, Canada}

\begin{abstract}
  We present a framework for differentiable quantum transforms. Such transforms
  are metaprograms capable of manipulating quantum programs in a way that
  preserves their differentiability. We highlight their potential with a set of
  relevant examples across quantum computing (gradient computation, circuit
  compilation, and error mitigation), and implement them using the transform
  framework of PennyLane, a software library for differentiable quantum
  programming. In this framework, the transforms themselves are differentiable
  and can be parametrized and optimized, which opens up the possibility of improved quantum
  resource requirements across a spectrum of tasks.
\end{abstract}

\maketitle

\section{Introduction} 

Quantum machine learning (QML) is a rapidly-growing area of research with great
potential. Tools for designing QML algorithms and applications are increasingly
being incorporated into open-source quantum software
\cite{bergholm2020pennylane, Qiskit, tequila, tfq, yao}. Some of these integrate
with classical machine learning tools such as Autograd \cite{autograd}, PyTorch
\cite{pytorch}, TensorFlow \cite{tensorflow}, and JAX \cite{jax2018github}. The
core functionality provided by all of these libraries is automatic
differentiation (autodifferentiation) of mathematical functions, which is used
to compute the gradients required for model training. The capabilities provided
by these libraries have enabled significant advances to be made in classical
machine learning, allowing developers and practitioners to focus more on
architectures, data, and algorithms, rather than on the technical implementation
of computing derivatives.

Differentiation is a process that maps a function $f(x)$ to another function,
$g(x) = \nabla f(x)$, which computes its derivative. Differentiation can thus be
viewed as a function \emph{transform}. Some classical frameworks, such as JAX
\cite{jax2018github}, Dex \cite{paszke2021getting}, and functorch
\cite{functorch2021} make this notion explicit. For example, JAX is branded as a
library for transforming numerical functions, and its built-in transforms
include just-in-time compilation and automatic vectorization (\texttt{vmap}) in
addition to autodifferentiation. The system is extensible and allows users to
write their own transforms. Furthermore, these transforms preserve
differentiability of whatever they act upon.

There are many processes in quantum computing that rely on the idea of
transforming a \emph{quantum function}, most often represented by a quantum
circuit.  Such \emph{quantum transforms} take a circuit as input, and return a
new, modified circuit as output. This idea, and the underlying functional
programming elements, is perhaps most explicit in the Haskell-based language
Quipper \cite{green2013quipper}, which contains a built-in, user-extensible
system for monad transformers that can be applied to quantum circuits. Such
transforms are used most often in the context of quantum compilation or
transpilation, wherein gates in a circuit are rewritten in terms of other gates,
reordered, or otherwise optimized. Many multipurpose quantum software libraries such as Qiskit
\cite{Qiskit} and Cirq \cite{cirq}, and the compiler t$\vert$ket$\rangle$
\cite{sivarajah2020tket}, expose such optimization transforms to the
user. However, the concept of quantum transforms extends beyond simply turning
one quantum circuit into another. A variety of quantum computing tasks can be
expressed in the language of transforms, and more specifically, transforms that
can preserve the ability to compute quantum gradients automatically.

In this work we formalize and implement \emph{differentiable quantum
  transforms}, and discuss three specific applications: gradient computation,
quantum compilation, and noise characterization and mitigation. We outline the
formal definitions and key features of transforms in \autoref{subsec:formalism},
and give an overview of the relevant examples in
\autoref{subsec:applications-formal}. We then present in \autoref{sec:pennylane}
an implementation of an extensible transform system using the PennyLane quantum
software library \cite{bergholm2020pennylane}. We show how the aforementioned
applications can be implemented, and present explicit software examples that
demonstrate its advantages. We conclude with ideas for future research and
applications. Most notably, the transforms themselves are fully
differentiable, so they can be parametrized and trained. This enables us not
only to find optimal transform parameters, but gives us the potential to
\emph{learn new transforms}.

\section{Differentiable quantum transforms}
\label{sec:transforms}

\subsection{Formalism}
\label{subsec:formalism}

At a high level, a differentiable quantum transform is a composable function
that takes a differentiable quantum program as input, and returns one or more
differentiable quantum programs as output. Let $S$ be a \emph{quantum program}
\cite{ying2016foundations, zhu2020principles}. A quantum program may prepare the
state of one or more qubits, apply quantum operations (or other quantum
programs), and terminates with the measurement of one or more qubits. The
probabilistic nature of measurement in quantum mechanics means that the output
of a quantum program is non-deterministic in nature. In the context of the
programs and transforms we discuss here, we will often consider the program
output to be the expectation value of some observable taken in the
limit of an infinite number of such measurements (shots), so that the output is
reproducible.

Operations applied during the course a quantum program may be unitary
operations, or quantum channels. An element of the unitary group on $n$ qubits,
$\mathcal{U}(2^{n})$, is parametrized by $2^{2n}$ real values. Many quantum
channels (e.g., depolarization, amplitude damping, etc.) are also expressed in
terms of some real-valued parameter. Thus, we express a \emph{parametrized}
quantum program as $S(\mathbf{\sigma})$, where
$\mathbf{\sigma} = (\sigma_{1}, \ldots, \sigma_{m})$ are the parameters
\cite{zhu2020principles}. Quantum programs can be differentiated with respect to
their parameters; we denote this by $\frac{\partial S}{\partial \sigma_{i}}$,
where it is implicit that the function being differentiated is the mathematical
function implemented by the program $S$.

\begin{defn}
  Let $S$ be a quantum program with input parameters $\{\sigma_{i}\}$. $S$ is a \emph{differentiable quantum
    program} if $\frac{\partial S}{\partial \sigma_{i}}$ is
  defined for all $\sigma_{i}$.
\end{defn}

Many derivatives (including higher-order ones) with respect to program
parameters can be computed and evaluated on hardware. This generally involves
the use of \emph{parameter-shift rules}, which evaluate the quantum program at
different values of its parameters and compute a function of the output
\cite{mitarai2018quantum, li2017hybrid, banchi2021measuring,
  schuld2019evaluating, wierichs2021general}. For example, the common two-term
shift rule which applies to many single-parameter gates is
\begin{equation}
 \frac{\partial S}{\partial \sigma_{i}} = c \left[ S(a \sigma_{i} + s) - S(a
   \sigma_{i} - s) \right],
\end{equation}

\noindent where $s$ is the shift parameter, and $a$ and $c$ are gate-dependent
constants (by default, $c = 1/2$, and $a = 1$).  While this appears similar to a
finite-difference method, the value of $s$ is macroscopic (typically $\pi/2$),
enabling gradients to be estimated even in a noisy hardware setting where the
infinitesimal shift of finite-differences would be completely washed out due to
noise.

A quantum transform $\mathcal{T}$ is a \emph{metaprogram} that deterministically
maps an input quantum program to one or more output quantum programs, and
optionally depends on one or more parameters $\{\tau_{i}\}$. A
\emph{differentiable quantum transform} (DQT) is a transform that preserves
differentiability of the input program with respect to the program parameters,
while itself being a differentiable program. As is the case with quantum
programs, we denote differentiation of a transform by
$\frac{\partial \mathcal{T}}{\partial \tau_{i}}$, where it is implicit that
the function being differentiated is the mathematical function implemented by $S$
after it is transformed by $\mathcal{T}$.

\begin{defn}
  Let $S$ be a differentiable quantum program with inputs $\{\sigma_{i}\}$. A
  program $\mathcal{T}$ with inputs $\{\tau_{i}\}$ is a \emph{differentiable
    quantum transform} if it maps $S$ to one or more
  output programs, i.e., 
  \begin{equation}
    \mathcal{T}(S) \rightarrow \{S_{k}^{\prime} \}
  \end{equation}
  where each $S_{k}^{\prime}$ is also a differentiable quantum program with
  respect to the same inputs $\{\sigma_{i}\}$, and
  $\frac{\partial \mathcal{T}}{\partial \tau_{i}}$ is defined for all
  $\{\tau_{i}\}$.
\end{defn}

\noindent Transforms for which $\{\tau_{i}\} = \emptyset$ are
\emph{non-parametrized transforms}, whereas those with
$\{\tau_{i}\} \neq \emptyset$ are \emph{parametrized transforms}. We denote
transforms mapping one program to a single other, $|\{S^{\prime}_{k}\}| = 1$, as
\emph{single transforms}. Transforms mapping one program to many,
$|\{S^{\prime}_{k}\}| > 1$, are termed \emph{batch transforms}. Differentiating
the output of a batch transform may consist of differentiating the output of
each program independently, or a function of those outputs.

We define both single and batch transforms to be mappable over lists of quantum
programs, i.e., for single transforms,
$\mathcal{T}([S_{1}, \ldots, S_{k}]) = [\mathcal{T}(S_{1}), \ldots,
\mathcal{T}(S_{k})]$, where in the case of a batch transform, the result is a
list of lists of programs. As such, transforms (whether parametrized or not) are composable.

\begin{corollary}
  Let $\mathcal{T}$, $\mathcal{U}$ be two DQTs. DQTs are \emph{composable}, i.e.,
  $\mathcal{V} = \mathcal{U} \cdot \mathcal{T}$ is also a DQT.
\end{corollary}

\noindent Composability of transforms enables us to construct and apply extensive
pipelines of transforms to quantum programs, all the while preserving the differentiability of that
program's input parameters.

\subsection{Applications}
 \label{subsec:applications-formal}

 \subsubsection{Gradient computation as transforms}
 \label{subsubsec:gradients-theory}
 
Computation of quantum gradients is a key application of batch
transforms. Consider, for example, the parameter-shift rules for
computing gradients of parametrized unitaries on quantum hardware.
Given a parametrized quantum circuit function
\begin{align}
  f(\theta) = \langle \psi \vert U(\theta)^\dagger \hat{B} U(\theta) \vert \psi \rangle,
\end{align}
where
\begin{itemize}
  \item $|\psi\rangle$ is the initial quantum state,
  \item $U(\theta)=e^{iG\theta}$ is some parametrized unitary, with generator $G$ having equidistant eigenvalues,
  \item and $\hat{B}$ is an observable we wish to compute the expectation value of,
\end{itemize}
the parameter-shift rule allows us to compute the partial derivative of the expectation
value by executing the same quantum function with (equidistant) shifted values \cite{wierichs2021general}:
\begin{align}
  \label{eq:gen-parameter-shift}
  \frac{\partial}{\partial \theta}f(\theta) =
  \sum_{\mu=1}^{2R} f\left(x + \frac{2\mu-1}{2R}\pi\right)
  \frac{(-1)^{\mu-1}}{4R\sin^2\left(\frac{2\mu-1}{4R}\pi\right)}.
\end{align}
Here, $R$ is the set of all (unique) pairwise differences of the eigenvalue spectrum of unitary
generator $G$\footnote{Note that since $G$ has equidistant eigenvalues, the set
  $R$ simply consists of some base frequency multiplied by natural numbers.}. In essence, we can evaluate the gradient of $f(\theta)$ with respect to a particular
parameter value $\theta$ by evaluating the circuit at $2R$ points, and then algebraically combining
the results. This could be accomplished by a batch transform which
creates a separate quantum program for each term in the sum, and then
combines the results according to \autoref{eq:gen-parameter-shift}.

Note that for the case where $R=1$ (corresponding to single-qubit rotation gates), the
above parameter-shift rule reduces to the commonly known two-term rule \cite{schuld2019evaluating}:
\begin{equation}
  \label{eq:parameter-shift}
  \frac{\partial U(\theta)}{\partial \theta} = \frac{1}{2} \left[
    U\left(\theta + \frac{\pi}{2}\right) - U\left(\theta - \frac{\pi}{2}\right) \right].
\end{equation}
This can be easily extrapolated
to other types of gradient computation, such as finite-differences, or for gates
that permit more complex parameter-shift rules (such as the controlled-rotation
gates with four-term shift rules).

 \subsubsection{Differentiable quantum compilation}

Quantum compilation is the process of decomposing a high-level specification of
a quantum algorithm into a sequence of elementary operations in a format
suitable for a particular quantum device. This is an extensive pipeline with
numerous components: circuit synthesis, circuit optimization, transpilation,
hardware-specific qubit placement and routing, gate scheduling, and optimization of
low-level pulse controls.

All these tasks can be naturally viewed as quantum transforms: a circuit is fed
in as input, and a modified and/or optimized circuit comes out. Compilation is
implemented in this manner in many other quantum software libraries. These
often consist of a pipeline of one or more \emph{passes} through a set of
subroutines that transform the circuit. Such subroutines typically manipulate
the circuit at the level of its directed acyclic graph (DAG). There is some
existing work which provides guidance on how to put together automated pipelines
\cite{nam2018automated}. Software tools for compilation such as t$|$ket$\rangle$
\cite{sivarajah2020tket}, \texttt{staq} \cite{staq}, \texttt{quilc}
\cite{quilc}, Qiskit \cite{Qiskit}, Cirq \cite{cirq}, XACC \cite{xacc}, QCOR
\cite{qcor1, qcor2}, Quipper \cite{green2013quipper},
and the transforms in PennyLane, all give the user the flexibility to define
pipelines based on a set of available building blocks. In particular,
t$|$ket$\rangle$ explicitly uses a modular transform system in its
software. Both t$|$ket$\rangle$ and pyquil have support for partial or
parametric compilation respectively, which makes compilation of parametrized
circuits more efficient by first tracing through execution with symbolic
variables, and then specifying the numerical parameter values at
runtime.

Implementing compilation using differentiable transforms allows for optimization
of parametrized circuits without compromising the differentiability of the
parameters. For example, consider a single transform that modifies a quantum
program by finding adjacent rotations of the same type on the same qubit, and
combining them into a single rotation, e.g.,
$RZ(\phi) RZ(\theta) = RZ(\phi + \theta)$. If both $\phi$ and $\theta$ are
trainable parameters, the sum of these values represents a new trainable
parameter, which we call $\lambda = \phi + \theta$. If implemented in an
autodifferentiable manner where operations on parameters are traced during a
forward pass, we can compute the gradient of $\lambda$ alone and use it to extract
the gradients of $\phi$ and $\theta$. This can be advantageous if using, for
example, the parameter-shift rule of \autoref{eq:parameter-shift}. With only one
trainable parameter, we must evaluate the program only twice, rather than four
times in the case where each gradient must be computed individually.

\subsubsection{Transforms and noise}

Many common tasks in both noise characterization and error mitigation can be
framed and explored in the context of differentiable transforms. The most basic
application is to use transforms to \emph{add} noise; a single parametrized
transform may modify a quantum program by inserting applications of a parametrized
noise channel after certain types of gates in order to simulate the behaviour
of a noisy device. As transforms are composable, such noise models are highly
customizable and allow for varying types and amounts of noise to be added.

However, perhaps a more interesting task is to leverage differentiable
transforms to \emph{characterize} noise. One can create a parametrized noise
model as described above, and use the results of experiments on a noisy device
to learn the values of the noise parameters that most closely match the observed
behaviour. Thanks to the autodifferentiability of the transform parameters, this
can be done using standard optimization techniques such as gradient descent.

Error mitigation methods are essential when running computations on noisy
near-term quantum devices. One such method, zero-noise extrapolation (ZNE), is a
technique that estimates a noiseless value of a result by running a circuit
for increasing values of some scale factor that adds noise, and then
extrapolating the results back to the zero-noise case \cite{li2017efficient,
  temme2017error, kandala2019error, mitiq}. ZNE naturally incorporates both
single and batch transforms: a number of common methods used for addition of
noise, such as CNOT pair insertion and unitary folding
\cite{giurgicatiron2020digital}, can be implemented as single transforms; we can
then implement a batch transform which uses the single transform to create new
programs with different amounts of noise, and then compute a function of the
results to obtain the mitigated value.

\section{Differentiable quantum transforms in PennyLane}
\label{sec:pennylane}

Going beyond the theoretical description, we will discuss and showcase quantum
transforms in the context of PennyLane. PennyLane, while not a pure functional
library, contains a significant proportion of functional elements which
enabled the development of the transforms module, \texttt{qml.transforms}
(\texttt{qml} is the standard import alias for PennyLane). We
begin with an overview of the key aspects of the system, followed by
implementations of the applications discussed in \autoref{subsec:applications-formal}.

\subsection{The \texttt{transforms} module}

\begin{figure}[htbp]
  \centerline{\includegraphics[]{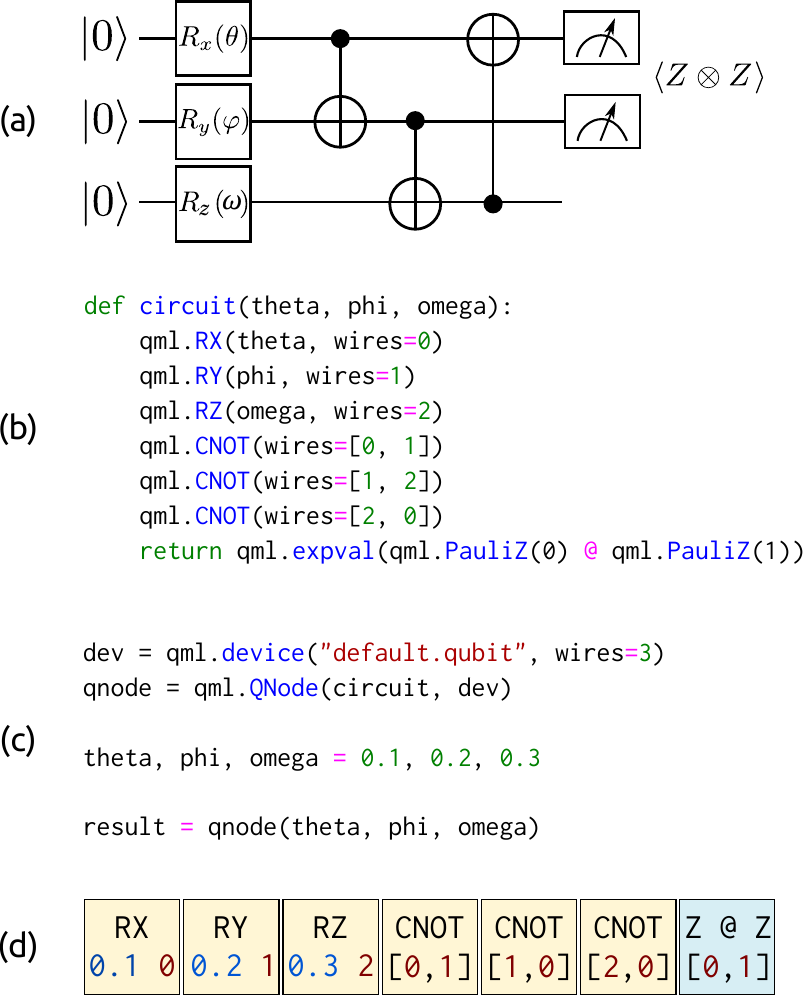}}
  \caption[]{Representations of quantum programs in PennyLane. (a) A standard quantum
    circuit. (b) A quantum function. (c) A quantum node (QNode), consisting of a
    quantum function bound to a quantum device on which it can be executed. (d)
    A quantum tape, a low-level data structure representing the quantum
    circuit which is constructed by the QNode and then
    executed on a device.\label{fig:data_structures} }
\end{figure}

PennyLane represents (differentiable) quantum programs and quantum circuits
using three different types of data structures: quantum functions, quantum nodes,
and quantum tapes, as shown in \autoref{fig:data_structures}. 
The core component of computation is a quantum function, which is a programmatic
representation of a quantum circuit. Quantum functions are regular Python
functions that accept arguments as input, apply a sequence of quantum
operations, and return one or more quantum measurements.

In order to run a quantum function and obtain measurement results, a quantum
function must be bound to a \emph{device}. Such a binding is accomplished using
a higher-level data structure called a \emph{quantum node}, or \emph{QNode}. The
device may be a simulator, or actual quantum hardware. Once the two are bound,
the quantum circuit can be executed by specifying the set of input parameters,
and calling the QNode using the same syntax and parameters as one would call the
underlying quantum function (this is enabled by the fact that the QNode wrapper
is actually a Python \emph{decorator}).

In order to actually run a QNode with a set of provided parameters, upon
invocation, the QNode constructs an internal representation of the quantum
function called a \emph{quantum tape}. A quantum tape is the lowest-level data
structure, representing a quantum program as an \emph{annotated queue} of
operations and measurements. Parameter values are assigned upon construction of
the tape.

PennyLane contains explicit construction mechanisms for single transforms, batch
transforms, and other types of non-composable transforms that can, e.g., be
applied to a QNode to extract information about it. In many cases, the
transforms can be applied to more than one type of data structure (for example,
the \texttt{qml.transforms.insert} transform can be applied to quantum
functions, QNodes, or even quantum devices, in order to insert gates at
specified positions in a quantum circuit). 

\subsubsection{Single-tape and quantum function transforms}

\begin{figure}[htbp]
  \centerline{\includegraphics[]{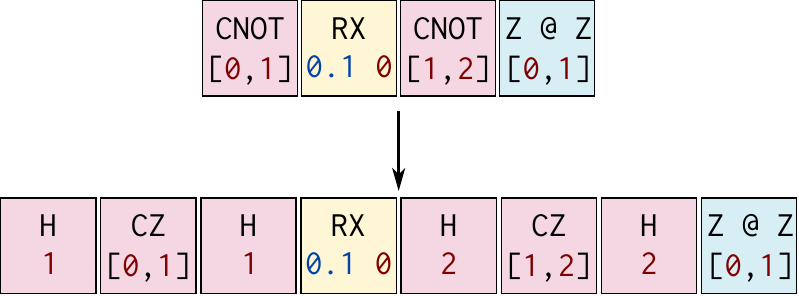}}
  \caption[]{Example of a quantum tape transform. This transform applies the
    circuit identity $CNOT_{ij} = H_{j} \cdot CZ_{ij}\cdot H_{j}$ to all instances of
    a $CNOT$ from any control qubit $i$ to target qubit $j$ on a quantum tape.  \label{fig:tape_transform} }
\end{figure}

Single-tape transforms are the base unit of transforms in PennyLane. These are
one-to-one transforms in which elements of a tape may be removed, added, or
modified. Furthermore, the transform may accept one or more parameters which
affect how the tape is modified. A simple example is presented in
\autoref{fig:tape_transform}, wherein all $CNOT$ gates are converted to $CZ$ and
Hadamard gates by way of a textbook circuit identity.

PennyLane tapes contain two lists: one of \texttt{operations}, and one of
\texttt{measurements}. The software implementation of a tape transform consists
of simply looping through the operations, and in the current context, queuing a
new set of modified operations. PennyLane provides convenience decorators to
ease the construction of tape transforms and other types of transforms.  For example, the
transform depicted in \autoref{fig:tape_transform} would be implemented in
PennyLane as follows:
 
 \begin{lstlisting}[language=python]
import pennylane as qml

@qml.single_tape_transform
def convert_cnots(tape):
    for op in tape.operations + tape.measurements:
        if op.name == 'CNOT':
            qml.Hadamard(wires=op.wires[1])
            qml.CZ(wires=[op.wires[0], op.wires[1]])
            qml.Hadamard(wires=op.wires[1])
        else: 
            qml.apply(op)
 \end{lstlisting}

One way to invoke the transform is by applying it to a quantum tape directly.
 
\begin{lstlisting}[language=python]
with qml.tape.QuantumTape() as tape:
    qml.CNOT(wires=[0, 1])
    qml.RX(0.1, wires=0)
    qml.CNOT(wires=[1, 2])
    qml.expval(qml.PauliZ(0) @ qml.PauliZ(1))

transformed_tape = convert_cnots(tape)
\end{lstlisting}

Quantum function transforms (qfunc transforms) are elevated tape transforms, and
simply wrap that underlying structure. Quantum tapes themselves are a
lower-level data structure, and generally accessed and manipulated internally
(e.g., constructed by a QNode prior to execution), rather than being
user-facing. Therefore, qfunc transforms enable the same functionality for a
user while still being able to work at the abstraction level of functions. They
are implemented in essentially the same way, and in fact a tape transform can be
converted to a qfunc transform simply by swapping out the top-level
decorator.

\begin{lstlisting}[language=python]
@qml.qfunc_transform
def convert_cnots(tape):
    for op in tape.operations + tape.measurements:
        if op.name == 'CNOT':
            qml.Hadamard(wires=op.wires[1])
            qml.CZ(wires=[op.wires[0], op.wires[1]])
            qml.Hadamard(wires=op.wires[1])
        else:
            qml.apply(op)
\end{lstlisting}

The advantage, now, is that the \texttt{convert\_cnots} transform can be
applied directly to quantum functions as a decorator, with no need to consider 
tapes at all beyond the implementation of the transform itself.

\begin{lstlisting}[language=python]
@convert_cnots
def circuit(param):
    qml.CNOT(wires=[0, 1])
    qml.RX(param, wires=0)
    qml.CNOT(wires=[1, 2])
    return qml.expval(qml.PauliZ(0) @ qml.PauliZ(1))
\end{lstlisting}

\noindent This flexibility enables us to easily \emph{compose} single qfunc transforms
by chaining functions, or by stacking decorators.

Transforms may also contain classical processing which affects the parameter
values of operations on tapes. PennyLane includes a special module, \texttt{pennylane.math},
which enables manipulation of the parameter values (more generally, tensors) in
a way that is agnostic to the underlying classical machine learning framework,
and thus preserves differentiability.  For example, we can write a single
transform that acts on all \texttt{qml.RX} rotations and rotates by the square
root of the original parameter value:

\begin{lstlisting}[language=python]
import pennylane.math as math

@qml.qfunc_transform
def square_root_rx(tape):
    for op in tape.operations + tape.measurements:
        if op.name == 'RX':
            qml.RX(math.sqrt(op.data[0]), wires=op.wires[0])
        else:
            qml.apply(op)
\end{lstlisting}

We can create a quantum function that applies \texttt{qml.RX} rotations, apply the
transform, and then compute the gradients with respect to the input
parameters in any framework. Below is an example using the PyTorch framework which illustrates
this flexibility.

\begin{lstlisting}[language=python]
import torch

def apply_rx(x):
    qml.RX(x, wires=0)
    return qml.expval(qml.PauliZ(0))

dev = qml.device('default.qubit', wires=1)

qnode = qml.QNode(apply_rx, dev, interface='torch')

x_orig = torch.tensor(0.3, requires_grad=True)
res = qnode(x_orig)
res.backward()

transformed_qnode = qml.QNode(square_root_rx(apply_rx), dev, interface='torch')

x_transformed = torch.tensor(0.3, requires_grad=True)
res = transformed_qnode(x_transformed)
res.backward()
\end{lstlisting}

\begin{lstlisting}[language=python]
>>> x_orig.grad
tensor(-0.2955)

>>> x_transformed.grad
tensor(-0.4754)
\end{lstlisting}

\subsubsection{Batch transforms}

\begin{figure}[htbp]
  \centerline{\includegraphics[]{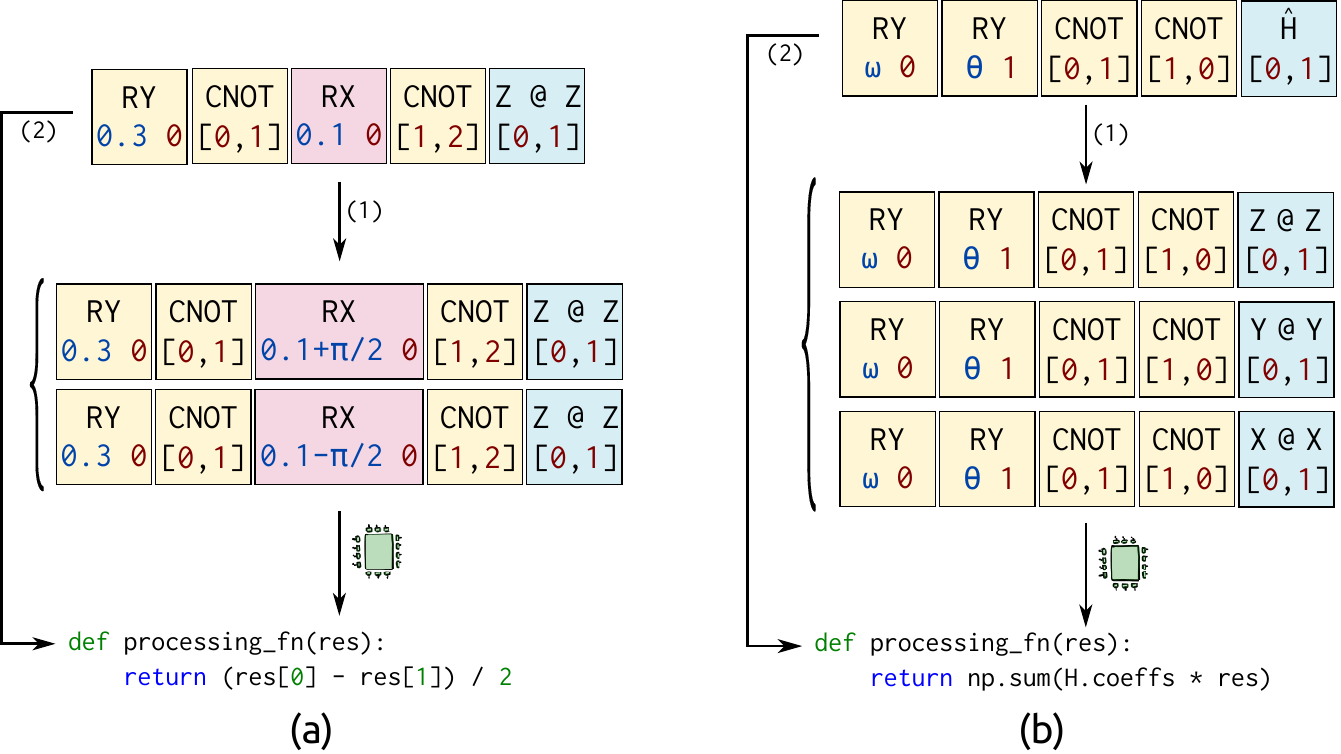}}
  \caption[]{
Visual depiction of two use cases of batch transforms. (a) 
    Given an input tape, a gradient batch transform returns two objects: (1) one
    or more new, transformed tapes, and (2) a classical function that accepts the
    \emph{results} of the executed transformed tapes, and returns a processed
    value. This particular example computes the gradient of the second parameter
    on the input tape using the parameter-shift rule of
    \autoref{eq:parameter-shift}. (b) A batch transform being used to evaluate the
    expectation value of a Hamiltonian $\hat{H} = c_{1} ZZ + c_{2} YY +
    c_{3}XX$. (1) For each Pauli term (or, group of
    commuting terms) in the Hamiltonian, a separate tape is created and
    executed, returning that term's expectation value. (2) The final expectation
    value, which is a linear combination of the Hamiltonian coefficients and
    associated expectation values, is computed by the processing function
    according to \autoref{eq:ham-exp-val} based
    on the execution results.}  \label{fig:batch_transform}
\end{figure}

Batch transforms are one-to-many transforms which take one tape as input, and
return a \emph{collection} of tapes as output. Furthermore, in PennyLane, they
may also return a classical processing function that acts on the results of the
executed quantum tapes to compute a desired quantity (this function should also
be differentiable). A key use case of batch transforms is the computation of
quantum gradients, which was discussed in \autoref{subsubsec:gradients-theory}
and is shown graphically in \autoref{fig:batch_transform}. The ability to
differentiate and compose transforms ensures that we can compute $n$-th order
derivatives without any obstacles.

Batch transforms can also be used to compute the expectation value of a
Hamiltonian, also depicted in \autoref{fig:batch_transform}. Let
\begin{equation}
\label{eq:1}
\hat{H} = \sum_{i} c_{i} P_{i}, \quad P_{i} \in \mathcal{P}_{n},
\end{equation}
where $\mathcal{P}_{n}$ is the $n$-qubit Pauli group, and let $U(\theta)$ be a
parametrized quantum circuit. We can compute the expectation value of $\hat{H}$,
$\langle \hat{H} \rangle$, after applying $U(\theta)$ as a linear combination of the
expectation value of all of its terms,
\begin{eqnarray}
  \langle \hat{H} \rangle &=& \sum_{i} c_{i} \bra{0} U^{\dag}(\theta)P_{i}U(\theta)
      \ket{0}. \label{eq:ham-exp-val}
\end{eqnarray}

\noindent To compute $\langle \hat{H} \rangle$ using a batch transform, we first transform
the initial tape into multiple tapes, each of which measures the expectation
value of an individual $P_{i}$ (alternatively, we can partition the terms of $\hat{H}$
into commuting sets, and compute one expectation value per set). We then execute
each of the tapes, and feed the results into a processing function which will
evaluate \autoref{eq:ham-exp-val}. Below we construct the Hamiltonian and tape
depicted in \autoref{fig:batch_transform}.

\begin{lstlisting}[language=python]
# Hamiltonian in Figure 3 where c_1 = c_2 = c_3 = 1
coeffs = [1, 1, 1]
obs = [
    qml.PauliZ(0) @ qml.PauliZ(1), 
    qml.PauliY(0) @ qml.PauliY(1), 
    qml.PauliX(0) @ qml.PauliX(1)
]

H = qml.Hamiltonian(coeffs, obs)

with qml.tape.QuantumTape() as tape:
    qml.RY(0.3, wires=0)
    qml.RY(0.4, wires=1)
    qml.CNOT(wires=[0, 1])
    qml.CNOT(wires=[1, 0])
    qml.expval(H)
\end{lstlisting}

We can now apply the batch transform (which is built-in to
PennyLane as \texttt{qml.transforms.hamiltonian\_expand}), and use the processing function
on the executed results.

\begin{lstlisting}[language=python]
tapes, fn = qml.transforms.hamiltonian_expand(tape)

dev = qml.device('default.qubit', wires=2)
res = dev.execute(tape)
\end{lstlisting}

\begin{lstlisting}[language=python]
>>> fn(res)
array([0.97272928])
\end{lstlisting}

Batch transforms can also be applied as a decorator directly to QNodes; the
outcome of executing the QNode is simply the output of the classical processing
function. If both the transform and processing function preserve
differentiability of all the parameters, the output of this processing function
can be fed as input into other parts of a differentiable quantum program, or
itself be differentiated (as will be demonstrated in \autoref{subsubsec:zne}).

\subsubsection{Other transforms}

The \texttt{qml.transforms} module contains two other types of transforms that
do not fit the criteria above: device transforms, and information
transforms. Device transforms act on PennyLane devices, modify their internal
workings, and then return a new device with different behaviour. An example is
\texttt{qml.transforms.insert}, which can add additional gates at specified
points in the circuit, e.g., to add simulated noise at the level of the device.

\begin{lstlisting}[language=python]
def circuit(x):
    qml.RX(-2*x, wires=0)
    qml.S(wires=0)    
    return qml.probs(wires=0)

dev = qml.device('default.mixed', wires=1)
qnode = qml.QNode(circuit, dev)

# Adds amplitude damping after every gate
noisy_dev = qml.transforms.insert(qml.AmplitudeDamping, 0.05, position="all")(dev)
noisy_qnode = qml.QNode(circuit, noisy_dev)
\end{lstlisting}

\begin{lstlisting}[language=python]
>>> print(qml.draw(qnode, expansion_strategy="device")(0.3))
 0: --RX(-0.6)--S--| Probs 
>>> print(qml.draw(noisy_qnode, expansion_strategy="device")(0.3))
 0: --RX(-0.6)--AmplitudeDamping(0.05)--S--AmplitudeDamping(0.05)--| Probs 
\end{lstlisting}

Information transforms are non-composable, non-differentiable transforms that take a tape, quantum
function, or QNode as input, and return a function capable of computing and/or
displaying information about that input. Key examples are \texttt{qml.draw},
used in the previous example, and \texttt{qml.specs}, which takes as input a
QNode and returns a function that computes its quantum resources.

\subsection{Examples}
\label{sec:examples}

In this section, we present implementations of the three examples of
\autoref{subsec:applications-formal} using the differentiable transforms
implemented in PennyLane. For each example, we detail a specific scenario in
which the differentiability yields significant advantages, insights, or enables
novel functionality. Unless otherwise noted, all examples below can be
implemented in the most recent release of PennyLane (v0.21).

 \subsubsection{Optimizing gradient computation in a noisy setting}

While the parameter-shift rule works for a large variety of gates in variational
quantum algorithms, we occasionally chance upon unitaries that we wish to train
on hardware that do not permit a parameter-shift rule. This could be for a variety
of reasons; perhaps the unitary does not satisfy the form $e^{iGx}$, or the
eigenvalue spectrum of its generator is unknown.

In such cases, we typically must fall back to numerical methods of differentiation
on hardware such as the method of finite-differences. Previous work exploring
finite-differences in a noisy setting has shown that the optimal finite-difference
step size for first-order forward difference is of the form \cite{mari2021estimating}
\begin{equation}
    \label{eq:optimum-step}
  h^* = \left(\frac{2 \sigma_0^2}{N f''(x)^2}\right) ^ {1 / 4},
\end{equation}
where $N$ is the number of shots (samples) used to estimate expectation values,
$\sigma_0$ is the single-shot variance of the estimates, and $f''(x)$ is the
second derivative of the quantum function at the evaluation point. While
for large $N$ we can make the approximation $h^*\approx N^{-0.25}$, for small $N$
on hardware, we must manually compute the second derivative of the quantum function in order
to determine a decent estimate for the gradient step-size, which can further introduce
error while adding a prohibitive number of additional quantum evaluations
required per optimization step. Instead, we can wrap the gradient computation in a quantum transform that learns
optimal parameters for the finite-difference step size in the presence of noise.

Consider the following variational quantum circuit:

\begin{lstlisting}[language=python]
N = 1000
dev = qml.device("default.qubit", wires=2, shots=N)

@qml.qnode(dev, max_diff=2)
def circuit(x):
    qml.Hadamard(wires=0)
    qml.Hadamard(wires=1)
    qml.SingleExcitation(x, wires=[0, 1])
    H = qml.PauliX(0) @ qml.PauliX(1)
    return qml.expval(H)
\end{lstlisting}

Here, \texttt{circuit} is the cost function we would like to optimize on a noisy
device using first-order forward finite-differences, and 1000 shots.
Rather than hard-coding in a constant finite-difference step size, we can include
the \text{variance} of the single-shot gradient as a quantity to minimize
in the cost function by using the \texttt{qml.gradients.finite\_diff}
quantum transform:

\begin{lstlisting}[language=python]
def cost_and_grad(x, h):
    """Return the cost function to minimize, and the quantum gradient"""
    g1 = qml.gradients.finite_diff(circuit, h=h)(x, shots=[(1, N)])
    return circuit(x) + np.var(g1) / N + h, np.mean(g1)

def cost(x, h):
    """Convenience function to return just the cost to minimize"""
    return cost_and_grad(x, h)[0]
\end{lstlisting}

Starting with $x=0.1$ and $h=10^{-7}$ (the default step size value of the \texttt{finite\_diff}
transform) we can now write an optimization loop that:

\begin{enumerate}
  \item Computes the cost value $f(x, h)$ and an estimate of the quantum gradient $\partial_x f(x,
        h)$ using single-shot finite differences with step-size $h$ (implemented
        together in \texttt{cost\_and\_grad}).
  \item Using autodifferentiation, computes the partial derivative of the cost value with respect to
        the step size, $\partial_h f(x, h)$.
  \item Applies a gradient descent step for both parameters $x$ and $h$.
\end{enumerate}

\begin{lstlisting}[language=python]
opt = qml.GradientDescentOptimizer(stepsize=0.05)

# PennyLane contains a wrapped version of NumPy which allows 
# for specification of trainable parameters using requires_grad
h = np.array(1e-7, requires_grad=True)
x = np.array(0.1, requires_grad=True)

h_track = []
cost_track = []

for i in range(300):
    h = np.clip(h, 0, 5)
    x = np.clip(x, 0, 2 * np.pi)

    h_track.append(h)
    cost_track.append(circuit(x))

    # as the cost function depends on the gradient of
    # the circuit wrt x, we return it alongside the loss value to
    # avoid additional computations
    loss, x_grad = cost_and_grad(x, h)

    # compute the gradient of the cost function wrt h
    h_grad = qml.grad(cost, argnum=1)(x, h)
    x, h = opt.apply_grad([x_grad, h_grad], (x, h))
\end{lstlisting}

\begin{figure}[h!]
  \centerline{\includegraphics[width=0.5\textwidth]{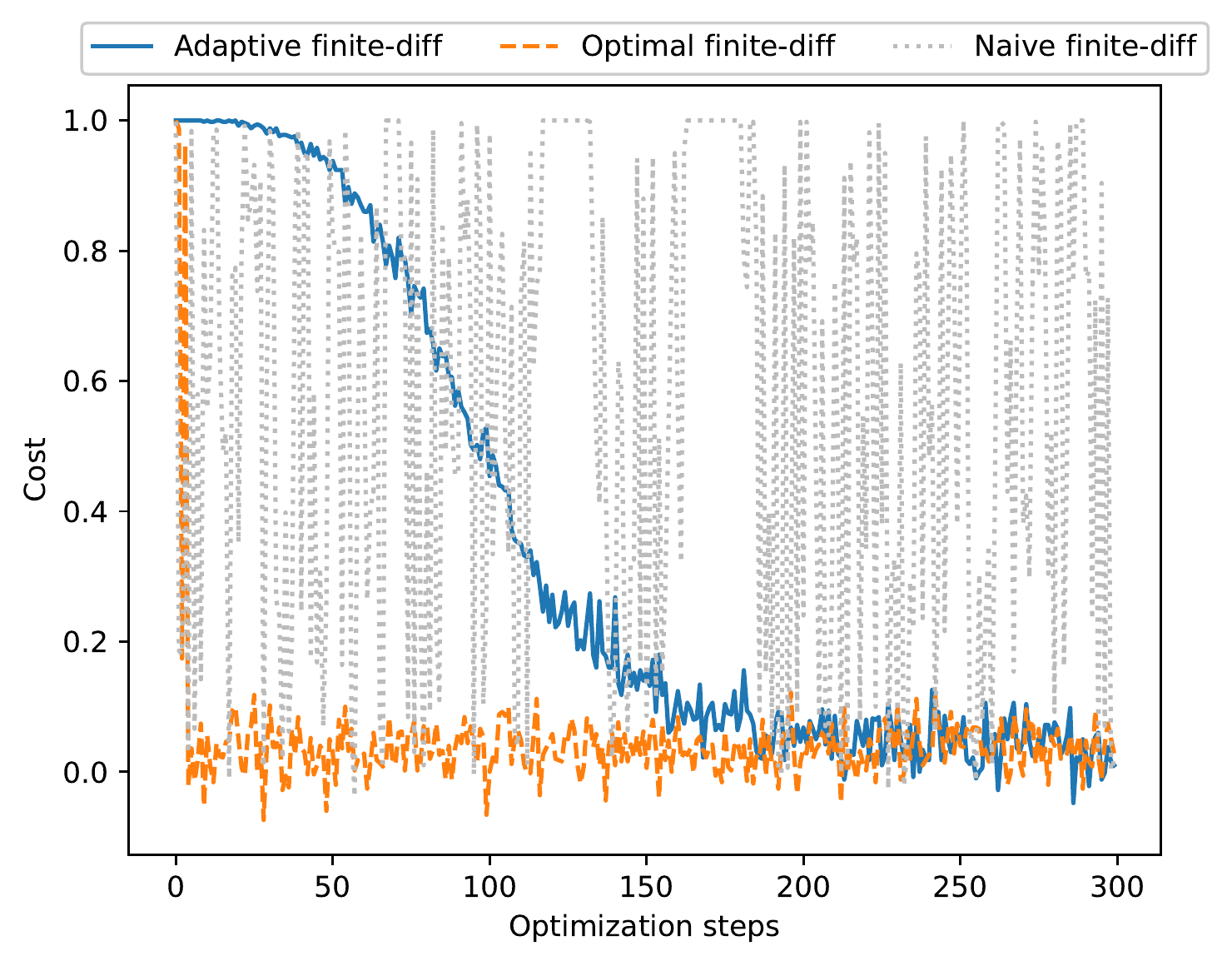}} \caption[]{Minimizing a
  two-qubit variational circuit cost function in a noisy setting, where expectation values must be
  approximated through sampling with $1000$ samples. Gradient descent is utilized for the
  optimization, with three different methods of computing the gradient at each optimization step:
  (a) first-order finite difference where the step size $h$ has been optimized alongside the cost
  function (blue, solid), (b) finite-difference optimization where the optimal step size $h^*$
  \autoref{eq:optimum-step} is computed at each iteration (orange, dashed), and (c) first-order
  finite differences with a constant step size of $h=10^{-7}$ (black, dotted).}
  \label{fig:adaptive-fd}
\end{figure}

The results of this `adaptive step-size finite-difference' optimization is compared to
both a  na\"ive  finite-difference optimization (using the default step size of $h=10^{-7}$)
and the optimal finite-difference optimization (by computing \autoref{eq:optimum-step} at every step)
in \autoref{fig:adaptive-fd}. It can be seen that the
na\"ive finite-difference optimization fails to converge to the minimum at all; a stepsize
of $10^{-7}$ results in a very large quantum gradient variance (as can be verified from
\autoref{eq:optimum-step}). The adaptive finite difference, as well as the
optimum finite difference,
by contrast, are both able to converge to the minimum --- the optimum variant converging particularly
quickly, as would be expected.

Nevertheless, the results show that knowing the underlying theoretical characteristics of a system
(such as in the optimal case) are not required. By simply encoding the quantity we wish to minimize
--- the expectation value \textit{and} the gradient variance --- the use of differentiable quantum
transforms allow us to train the model hyperparameters to minimize error during gradient descent.
Such approaches may also be viable in more complex models, where optimal or error-minimizing
hyperparameter values are not known in advance.

\subsubsection{Augmenting differentiable compilation transforms with JIT compilation}

The core difference between circuit compilation in PennyLane and other quantum
software libraries is its compilation routines are quantum
transforms, nearly all of which preserve differentiability\footnote{At the time
  of writing, the two-qubit unitary decomposition remains
  non-differentiable due to it involving non-differentiable
  library functions such as eigensystem computation.}. This enables the computation of
gradients of compiled circuits using a preferred autodifferentiation
framework. Furthermore, such gradient computation is also less resource
intensive because the autodifferentiation framework keeps track of the changes
in variables and can produce circuits with a reduced number of parameters.

All compilation transforms in PennyLane are implemented as
\texttt{qfunc\_transforms}, and as such, manipulate a circuit at the level of
its tape. Transforms include rotation merging, single-qubit gate fusion,
inverse cancellation, and moving single-qubit gates through control/target
qubits of controlled operations.

A top-level \texttt{qml.compile} transform is made available to the user to
facilitate creation of custom compilation pipelines. For example, the following
code shows a pipeline consisting of pushing commuting gates left through
controls and targets of two-qubit gates, and then fusing all sequences adjacent
of single-qubit gates into a single \texttt{qml.Rot} (general parametrized
unitary) operation.

\begin{lstlisting}[language=python]
pipeline = [
    qml.transforms.commute_controlled(direction='left'),
    qml.transforms.single_qubit_fusion
]

dev = qml.device('default.qubit', wires=3)

@qml.qnode(dev)
@qml.compile(pipeline=pipeline)
def circuit(x, y, z):
    qml.CNOT(wires=[0, 1])
    qml.RX(x, wires=1)
    qml.RY(y, wires=1)
    qml.S(wires=1)
    qml.CNOT(wires=[1, 2])
    qml.Hadamard(wires=2)
    qml.CNOT(wires=[2, 0])
    qml.RZ(z, wires=2)
    return qml.expval(qml.PauliZ(1))
\end{lstlisting}

\begin{figure}[htbp]
  \centerline{\includegraphics[width=0.5\columnwidth]{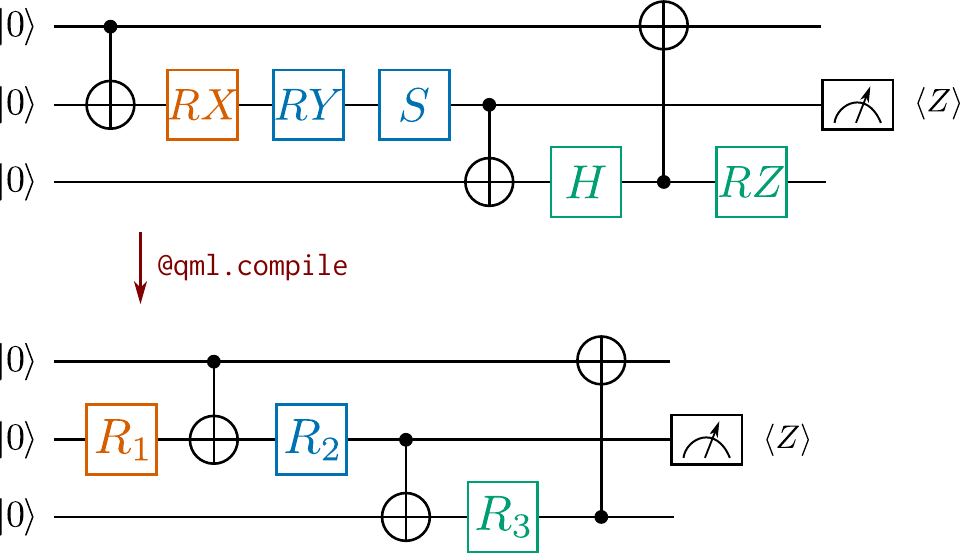}}
  \caption[]{Result of applying \texttt{qml.compile} with the transform pipeline
    \texttt{[commute\_controlled(direction='left'), single\_qubit\_fusion]} to a
    quantum circuit. Single-qubit gates are pushed left through controls and
    targets, and then fused into \texttt{qml.Rot} gates. Here,
    $R_{1} = \texttt{qml.Rot}(1.57, 0.1, -1.57)$,
    $R_{2} = \texttt{qml.Rot}(0, 0.2, 1.57)$, and
    $R_{3} = \texttt{qml.Rot}(3.14, 1.57, 0.3)$.}
  \label{fig:compiled_circuit}
\end{figure}

\autoref{fig:compiled_circuit} depicts the original and compiled circuit
obtained when running with input parameters \texttt{(0.1, 0.2,
  0.3)}. Furthermore, the compiled circuit remains fully differentiable with
respect to the input arguments, even though they may not appear directly as the
arguments in any of the gates of the compiled circuit.

\begin{lstlisting}[language=python]
>>> params = np.array([0.1, 0.2, 0.3], requires_grad=True)
>>> grad_fn = qml.grad(circuit)
>>> grad_fn(*params)
(array(-0.0978434), array(-0.19767681), array(1.33356867e-17))
\end{lstlisting}

A disadvantage of applying transforms in this way is that every circuit
execution involves feeding the quantum function through the transform pipeline. For large
circuits and multistep pipelines that involve more mathematically complex
operations such as full fusion of single-qubit gates, this could lead to
significant temporal overhead. This is especially undesirable for gradient
computations, as these by nature involve multiple executions of a quantum
circuit.

Most of PennyLane's circuit compilation transforms have been written in a way
that will remain differentiable even after applying just-in-time (JIT)
compilation\footnote{The two-qubit unitary decomposition is currently
  non-jittable due to use of conditional statements that determine the
  particular form of the decomposed circuit.}. This yields benefits in
both the classical and quantum aspects of running an algorithm. The classical
preprocessing, which involves applying the quantum transforms, becomes
significantly faster after the first jitted evaluation. We can then run the
optimized circuit on quantum devices, which will typically have lower depth and
fewer operations, without additional overhead. Furthermore, the same can be done
for computation of gradients: not only may the number of quantum evaluations be
reduced as a result of compiling the circuit, but the jitted gradient, after the
first execution, will run significantly faster than the original.

Here we  show an explicit example using the JAX
interface and \texttt{jax.jit}\footnote{The functionality required to enable
  jit with JAX is not available in v0.21 of PennyLane; it will be included in
  the next release. These results can be reproduced using the code on the open
  PR \#1894.}. Consider the following circuit:

\begin{lstlisting}[language=python]
def circuit(x, weights):
    for wire in range(5):
        qml.RX(x[wire], wires=wire)
        qml.Hadamard(wires=wire)
        qml.Rot(*weights[wire, :], wires=wire)
    
    for wire in range(5):
        qml.CNOT(wires=[wire, (wire + 1) % 5])
    
    return qml.expval(
        qml.PauliY(0) @ qml.PauliY(1) @ qml.PauliY(2) @ qml.PauliY(3) @ qml.PauliY(4)
    )
\end{lstlisting}

This circuit performs 3 layers of single-qubit gates, followed by a ring of
CNOTs. On a 5-qubit device, there are 20 total parameters. If we use
\texttt{jax.grad} to evaluate the gradient with the parameter-shift rules, we
expect to see 41 device executions (two per parameter, plus one for the initial
forward pass).

\begin{lstlisting}[language=python]
import jax
from jax import numpy as jnp

dev = qml.device('default.qubit', wires=5)

x = jnp.array([0.1, 0.2, 0.3, 0.4, 0.5])

weights = jnp.array([
    [-0.28371043,  0.93681631, -1.00500712],
    [ 1.41650132,  1.05433029,  0.91081303],
    [-0.42656701,  0.98618842, -0.55753227],
    [ 0.01532506, -2.07856628,  0.55483725],
    [ 0.91423682,  0.57445956,  0.72278638]]
)

original_qnode = qml.QNode(
    circuit, dev, interface="jax", diff_method="parameter-shift"
)

with qml.Tracker(dev) as tracker:
    jax.grad(original_qnode, argnums=(0, 1))(x, weights)
\end{lstlisting}

\begin{lstlisting}[language=python,mathescape=true]
>>> tracker.totals
{'executions': 41, 'batches': 2, 'batch_len': 41}
>>> %%timeit
... jax.grad(original_qnode, argnums=(0, 1))(x, weights)
870 ms $\pm$ 2.15 ms per loop (mean $\pm$ std. dev. of 7 runs, 1 loop each)
\end{lstlisting}

The time was obtained from running in a Jupyter notebook cell on an Intel i7 3.40GHz
processor. We can run the same circuit, but now compile using the
\texttt{qml.transforms.single\_qubit\_fusion} transform, which will merge all
adjacent single-qubit gates. This means our original 20-parameter circuit
now has effectively 15 parameters, as well as lower circuit depth (expanding the
rotations gives an original depth of 10, and compiled depth of 8).

\begin{lstlisting}[language=python]
dev = qml.device('default.qubit', wires=5)

compiled_qnode = qml.QNode(
    qml.transforms.single_qubit_fusion()(circuit), 
    dev, 
    interface="jax", 
    diff_method="parameter-shift"
)

with qml.Tracker(dev) as tracker:
    jax.grad(compiled_qnode, argnums=(0, 1))(x, weights)
\end{lstlisting}

\begin{lstlisting}[language=python,mathescape=true]
>>> tracker.totals
{'executions': 31, 'batches': 2, 'batch_len': 31}
>>> %%timeit
... jax.grad(compiled_qnode, argnums=(0, 1))(x, weights)
1.79 s $\pm$ 3.2 ms per loop (mean $\pm$ std. dev. of 7 runs, 1 loop each)
\end{lstlisting}

Even though the number of quantum evaluations is lower, the application of
transforms adds significant time overhead. But now, we can run \texttt{jit} on
both the original and compiled gradient functions to speed up the
computation. The first execution of the jitted function is typically longer
(in this case, roughly 0.4s for the original, and 16s for the compiled case),
however subsequent evaluations are markedly faster:

\begin{lstlisting}[language=python,mathescape=true]
jitted_original_grad = jax.jit(jax.grad(original_qnode, argnums=(0, 1)))
jitted_compiled_grad = jax.jit(jax.grad(compiled_qnode, argnums=(0, 1)))
\end{lstlisting}

\begin{lstlisting}[language=python,mathescape=true]
>>> jitted_original_grad(x, weights) # Run once
>>> %%timeit
... jitted_original_grad(x, weights)
24.7 ms $\pm$ 170 $\mu$s per loop (mean $\pm$ std. dev. of 7 runs, 10 loops each)
>>> jitted_compiled_grad(x, weights)
>>> %%timeit
... jitted_compiled_grad(x, weights)
14.4 ms $\pm$ 50.4 $\mu$s per loop (mean $\pm$ std. dev. of 7 runs, 100 loops each)
\end{lstlisting}

New values of the parameters can be passed to the jitted and compiled gradient
function without needing to run \texttt{jit} again. In variational algorithms,
where a circuit and its gradient are evaluated on the order of thousands of
times with different parameters over the course of the optimization process,
this would lead to a substantial speedup. While this is only a small example, it
demonstrates that ensuring compilation transforms preserve differentiability of
the input parameters can lead to benefits both when executing on a quantum
device and on a simulator.

\subsubsection{Differentiable error mitigation}
\label{subsubsec:zne}

In this section we will implement fully differentiable ZNE with linear
extrapolation (i.e., the extrapolated value is itself differentiable with
respect to the input circuit parameters). We demonstrate this using the
technique of unitary folding \cite{giurgicatiron2020digital}. In this method, a
circuit $U$ is first applied, followed by repetitions of
$U^{\dagger}U$. Following the example of the error-mitigation library
\texttt{mitiq} \cite{mitiq}, the number of such folds $n_{f}$ is computed based
on a scale factor $\lambda$ according to the expression
$n_{f} = (\lambda - 1)/2$, rounded to the nearest integer. A single transform
implementing this is shown below:

\begin{lstlisting}[language=python]
@qml.qfunc_transform
def unitary_folding(tape, scale_factor):
    for op in tape.operations:
        qml.apply(op)

    num_folds = math.round((scale_factor - 1.0) / 2.0)
    
    for _ in range(int(num_folds)):
        for op in tape.operations[::-1]:
            op.adjoint()

        for op in tape.operations:
            qml.apply(op)

    for m in tape.measurements:
        qml.apply(m)
\end{lstlisting}

ZNE fits perfectly into PennyLane's batch transform system. A batch transform
creates and returns multiple versions of the initial
tape with different amounts of folding, and returns a function \texttt{fit\_zne} which performs the noise extrapolation
on the results of executing those tapes. We perform the extrapolation by
hand-coding a simple linear regression (as we will soon demonstrate, we do so to
ensure that differentiability with respect to circuit input parameters is preserved).

 \begin{lstlisting}[language=python]
from pennylane.tape import stop_recording
from functools import partial

def fit_zne(scale_factors, energies):
    scale_factors = math.stack(scale_factors)
    unwrapped_energies = math.stack(energies).ravel()

    N = len(energies)

    sum_scales = math.sum(scale_factors)
    sum_energies = math.sum(unwrapped_energies)

    numerator = N * math.sum(
        math.multiply(scale_factors, unwrapped_energies)
    ) - sum_scales * sum_energies
    denominator = N * math.sum(scale_factors ** 2) - sum_scales ** 2
    slope = numerator / denominator

    intercept = (sum_energies - slope * sum_scales) / N

    return intercept


@qml.batch_transform
def zne(tape, mitigation_transform, scale_factors):
    with stop_recording():
        tapes = [mitigation_transform.tape_fn(tape, scale) for scale in scale_factors]

    processing_fn = partial(fit_zne, scale_factors)

    return tapes, processing_fn
 \end{lstlisting}

 The \texttt{zne} batch transform can now be applied either to a
 quantum tape to obtain the transformed tapes and processing function, or
  to a QNode to directly obtain the mitigated
 value. In fact once defined, a user can receive error-mitigated
 results simply by adding a single line to their code: the \texttt{@zne}
 decorator. This is demonstrated below, facilitated by the PennyLane-Qiskit
 plugin to import and apply a noise model to a device. We design a simple
 circuit which computes the expectation value of a multi-term Hamiltonian,
 as would be the case in near-term algorithms running on noisy devices, such as
 the variational eigensolver.
 
 \begin{lstlisting}[language=python]
from qiskit.test.mock import FakeVigo
from qiskit.providers.aer import QasmSimulator
from qiskit.providers.aer.noise import NoiseModel

device = QasmSimulator.from_backend(FakeVigo())
noise_model = NoiseModel.from_backend(device)
noisy_dev = qml.device(
    "qiskit.aer", backend='qasm_simulator', wires=3, shots=10000, noise_model=noise_model
)
noisy_dev.set_transpile_args(**{"optimization_level" : 0})

H = qml.Hamiltonian(
    coeffs=[1.0, 2.0, 3.0], 
    observables=[
        qml.PauliZ(0) @ qml.PauliZ(1), 
        qml.PauliZ(1) @ qml.PauliZ(2), 
        qml.PauliX(0) @ qml.PauliX(1) @ qml.PauliX(2)
    ]
)

@zne(unitary_folding, [1.0, 3.0, 5.0, 7.0, 9.0])
@qml.qnode(noisy_dev, diff_method='parameter-shift')
def circuit(params):
    qml.RX(params[0], wires=0)
    qml.CNOT(wires=[0, 1])
    qml.RY(params[1], wires=1)
    qml.CNOT(wires=[1, 2])
    qml.RZ(params[2], wires=2)
    qml.CNOT(wires=[2, 0])
    return qml.expval(H)

params = np.array([0.5, 0.1, -0.2], requires_grad=True)
\end{lstlisting}

\begin{lstlisting}[language=python]
>>> circuit(params)
2.9115300000000004
\end{lstlisting}
 
 Recall that we have programmed the linear regression manually using the
 \texttt{qml.math} module. This ensures that we can take the gradient of the mitigated value:
 
\begin{lstlisting}[language=python]
>>> qml.grad(circuit)(params)
array([-0.54936 ,  1.971385,  0.021725])
\end{lstlisting}
 
 In addition to using the \texttt{qml.transforms} module functionality to build
 ZNE methods from scratch, PennyLane includes a batch transform,
 \texttt{mitigate\_with\_zne}, that integrates directly with
 \texttt{mitiq}. However, computing gradients is not currently supported when
 using the \texttt{mitiq} backend.

 \subsubsection{Learning noise parameters with transforms}

 \label{subsubsec:characterization}
 
 The trainability of transform parameters can be leveraged for performing
 characterization tasks, such as learning parameters of noise. Suppose we
 have a simple noisy device where every single-qubit gate is depolarized by the
 same qubit-dependent amount. We can simulate this noise using a transform that
 applies an appropriate depolarization channel after every single-qubit gate.

\begin{lstlisting}[language=python]
@qml.qfunc_transform
def apply_depolarizing_noise(tape, p):
    for op in tape.operations:
        qml.apply(op)
        
        if len(op.wires) == 1:
            qml.DepolarizingChannel(p[int(op.wires[0])], wires=op.wires[0])
            
    for m in tape.measurements:
        qml.apply(m)
\end{lstlisting}

Suppose the set of true depolarization parameters is $\mathbf{p} = [0.05, 0.02]$, i.e.,
$p = 0.05$ for the first qubit, and $p = 0.02$ for the second. By choosing a
suitable circuit for experimentation, we can set up an optimization loop
that will use the transform to learn \emph{both} depolarization
parameters. For instance, let us create the following circuit:
 
\begin{lstlisting}[language=python]
def circuit(angles):
    qml.RX(angles[0], wires=0)
    qml.RY(angles[1], wires=1)
    return qml.expval(qml.PauliZ(0)), qml.expval(qml.PauliZ(1))
\end{lstlisting}

We first set up a representation of our noisy device, by creating a QNode and
applying the depolarization transform that uses the true depolarization
parameters. We set the device to use 10000 shots. Note that this step is
merely for simulation purposes, and can simply be replaced by a noisy device
that does not apply any transforms at all.

\begin{lstlisting}[language=python]
true_deps = np.array([0.05, 0.02])

noisy_dev = qml.device('default.mixed', wires=2, shots=10000)
noisy_circuit = apply_depolarizing_noise(true_deps)(circuit)

noisy_qnode = qml.QNode(noisy_circuit, noisy_dev)
\end{lstlisting}

Now, we construct an optimization loop to learn these depolarization parameters. We
use a simple least-squares loss for the cost, and compute the difference between
the output of a transformed QNode whose transform parameters we are trying to
learn (which is running on an ideal, simulated device), and the noisy QNode
(which is running on a device with the true amount of depolarization
noise). Note that here, we can evaluate the QNodes at arbitrary angular
parameters. We also must set initial values for the trainable parameters; we
choose something relatively high, and we will clip the values to between 0 and
1, to maintain validity of the values as parameters for a depolarization
channel.
 
\begin{lstlisting}[language=python]
angles = np.array([0.2, 0.3], requires_grad=False)
ideal_dev = qml.device('default.mixed', wires=2)

def cost(learn_eps):
    training_qnode = qml.QNode(apply_depolarizing_noise(learn_eps)(circuit), ideal_dev)
    return sum((training_qnode(angles) - noisy_qnode(angles)) ** 2)

opt = qml.GradientDescentOptimizer(stepsize=0.05)
correction_eps = np.array([0.1, 0.1], requires_grad=True)

np.random.seed(0) # For reproducibility

for _ in range(100):
    correction_eps = np.clip(correction_eps, 0, 1)
    correction_eps = opt.step(cost, correction_eps)
\end{lstlisting}

\begin{lstlisting}[language=python]
>>> print(correction_eps)
tensor([0.04973839, 0.02163015], requires_grad=True)
\end{lstlisting}

The learned parameters are close to the true values. While this is a simple
example, it could be generalized to more complex noise models, such as gate- and
qubit-dependent noise, other types of noise such as gate over-rotation, as well
as composition of noise channels.

 \section{Conclusions}

 Differentiable quantum transforms, and their implementation in PennyLane, are a
 flexible and highly-extensible tool for developing quantum algorithms. The
 examples shown here only scratch the surface of what transforms can enable
 in the future.

 For example, transforms for quantum gradients can be augmented to
 ensure that all hyperparameters themselves are trainable. Like we did
 for finite-differences, similar approaches can be applied to analytic
 gradient methods such as the parameter-shift formula and the Hadamard
 test. Without needing to delve into deep characterization, the gradient
 hyperparameters can be included in the cost function to be minimized,
 allowing the classical optimization loop to find optimal hyperparameter values
 to mitigate hardware error and improve convergence.

 Recent work \cite{ravi2021vaqem} demonstrated the advantage of a variational
 approach to error mitigation: this could be implemented and extended with
 transforms to perform similar tasks that can optimize parameters of an error
 mitigation protocol that uses a trainable transform concurrently with the
 optimization of parameters of a variational algorithm. The example in
 \autoref{subsubsec:characterization} in particular could be used as a starting
 point for exploration into data-driven approaches to characterizing devices in
 order to create adaptive, more realistic simulations of hardware. Such
 simulated devices could then be used as resources to learn recovery transforms
 to \emph{undo} noise, and develop improved error-mitigation techniques, without
 the overhead of needing to simulate on actual hardware.
 
 Finally, this unlocks, in all domains, the possibility for learning \emph{new}
 transforms. This would be of particular interest for applications in
 compilation and circuit optimization. Methods for, e.g., variational
 compilation \cite{khatri2019qaqc} and differentiable architecture search
 \cite{zhang2021differentiable} have been investigated, and parametrized
 transforms would enable similar methods to be seamlessly integrated into
 algorithmic pipelines.
 
 \section{Acknowledgements}

The authors thank Zeyue Niu, David Wierichs, and Richard Woloshyn for useful discussions.
 
\bibliography{main.bib}
\end{document}